\documentclass[12pt,prc,unsortedaddress,superscriptaddress,a4paper,nofootinbib,%
nobalancelastpage,preprintnumbers,showpacs,tightenlines
]{revtex4}
\usepackage{amsmath, amsbsy}
\usepackage{bbm}
\newif\ifpdf
        \ifx\pdfoutput\undefined
        \pdffalse 
        \else
        \pdfoutput=1 
        \pdftrue
        \fi
\ifpdf
        \usepackage[pdftex]{graphicx}
        \DeclareGraphicsExtensions{.pdf, .jpg}
\else
        \usepackage[dvips]{graphicx}
                \DeclareGraphicsExtensions{.eps, .jpg}
\fi

\def\vec#1{\boldsymbol{#1}}
\def\be{\begin{equation}}
\def\bel#1{\begin{equation}\label{#1}}
\def\ee{\end{equation}}
\begin{document}
\title{Stability of the pentaquark in a naive string model}
\author{Jean-Marc Richard}
\email{j-m.richard@ipnl.in2p3.fr}
\affiliation{Laboratoire de Physique Subatomique et Cosmologie,
 IN2P3-CNRS, Universit\'e Joseph Fourier,  INPG,
 53, avenue des Martyrs, 38026 Grenoble, France}
\affiliation{Institut de Physique Nucl\'eaire de Lyon,
Universit\'e de Lyon, IN2P3-CNRS, \\
43, rue du 11 Novembre 1918, 69622 Villeurbanne, France}
\date{\today}
\begin{abstract}
The  pentaquark is studied in a simple model of confinement where the quarks and the antiquark are linked by  flux tubes of minimal cumulated length, and the Coulomb-like interaction, the spin-dependent terms and the antisymmetrization constraints are neglected.. The ground-state is found to be stable against spontaneous dissociation into a meson and a baryon, both in the case of five equal-mass constituents and for a static quark or antiquark surrounded by four equal masses.
\end{abstract}
\maketitle

\section{Introduction}
In the 60s, some indications came up for a possible ``$Z$'' baryon resonance with  strangeness $S=+1$, but the data on the kaon--nucleon interaction did not confirm the existence of this state. For references, see early issues of the Review of Particle Properties, as cited in the last one~\cite{Amsler:2008zzb}. In the modern language, such a $Z$ resonance would be a $(\bar{s} nnnn)$ state, where $n$ denotes  a light quark $u$ or $d$.

In the 70s, some excitement arose on baryonium candidates, tentatively interpreted as $(nn-\bar{n}\bar{n})$ states with a separation between the $(nn)$ diquark and the $(\bar{n}\bar{n})$ antidiquark, and possibly an exotic color charge for these diquarks, see, e.g.,~\cite{Chan:1978nk}. It was then suggested that multiquark baryons could exist as well, with a structure $(q\bar{q}-qqq)$, and again, a possible orbital barrier between the clusters and perhaps a color-octet content of each cluster, this preventing from immediate rearrangement and subsequent decay into two color-singlet hadrons~\cite{deCrombrugghe:1978hi,Aerts:1977rw}. This work was abandoned when the evidence for baryonium faded away.

In 1977, Jaffe suggested that the dibaryon $H(uuddss)$ could be bound below the lowest threshold $\Lambda(uds)+\Lambda(uds)$ due to a coherence in the chromomagnetic interaction~\cite{Jaffe:1976yi}. In 1987, Lipkin~\cite{Lipkin:1987sk} and, independently, Gignoux et al.~\cite{Gignoux:1987cn}, pointed out that the same mechanism would bind a heavy pentaquark (the word was invented in these circumstances) such as $P=(\bar{c}uuds)$,  $(\bar{c}udds)$  or $(\bar{c}udss)$. This pentaquark was searched for in an experiment at Fermilab \cite{Aitala:1997ja}, which turned out not conclusive.  Further works indicated that the stability of $H$ and $P$ hardly survives a more consistent treatment of the short-range correlations which enter the chromomagnetic matrix elements and  the breaking of the SU(3) flavor symmetry in the light quark sector \cite{Oka:1983ku,Rosner:1985yh,Karl:1987uf}.

More recently, it was shown~\cite{Diakonov:1997mm} that in some  models of chiral dynamics, new baryons are predicted, in particular an antidecuplet  above the usual octet $(N,\, \Lambda,\ldots)$ and decuplet $(\Delta, \ldots, \Omega^-)$. This triggered a search by Nakano et al.\ (LEPS collaboration)~\cite{Nakano:2003qx}, who found  evidence for a baryon with strangeness $S=+1$. Much confusion followed, as stressed in~\cite{Amsler:2008zzb}, where  a critical review can be found, with  skeptical remarks not so much on the pioneering  theoretical speculation and experimental search, but on the followers. See, also, \cite{Tariq:2007ck}. Indeed, custom models were quickly designed, where the pentaquark was found with either positive and negative parity, made of ad-hoc quark clusters. Several experimentalists discovered the potential of their set-up and  stored data for looking at exotics and  hastily constructed mass spectra that could have been investigated much earlier. Eventually, experiments with high statistics and  good particle identification found no confirmation of the pentaquark candidates~\cite{Amsler:2008zzb}. There is now reasonable consensus that the light pentaquark does not exist, though some puzzling positive indications are still reported~\cite{Nakano:2008ee}.  For a recent discussion, see, e.g.,~\cite{Seth:2009dn}.

Nevertheless, the question of multiquarks remains important. On the experimental side, several states have been discovered in the hidden-charm sector~\cite{Amsler:2008zzb}, whose properties suggest large $(cq\bar{c}\bar{q})$ components, where $q$ is light or strange. On the theoretical sides,  multiquark states are now studied with the  QCD sum rules~\cite{Matheus:2006xi,Navarra:2007yw} and the lattice QCD \cite{Suganuma:2008ej}. In the past, the issue of multiquark states was mostly the field of constituent models: in principle, the basic ingredients can be tuned by fitting the spectrum of ordinary mesons and baryons, and then applied to tentative multiquark configurations. 
But the main difficulty lies in extrapolating the potential from the meson sector to larger systems. For the Coulomb-like part, in particular one-gluon exchange, the color additive rule
\begin{equation}\label{eq:col-add}
V=-\frac{3}{16}\,\sum_{i<j}\tilde\lambda^{(c)}_i.\tilde\lambda^{(c)}_j\,v(r_{ij})~,
\end{equation}
is probably justified. Here, $\tilde\lambda^{(c)}_i$ denotes the 8-vector color generator of the $i^\text{th}$ quark (with a suitable change for the antiquark) and the normalization is such that $v(r)$ holds for a color-singlet quark--antiquark meson.

However, there is no reason to use additive rule (\ref{eq:col-add}) for the confining part, although it has been often adopted as a tentative approximation. For baryons, a $Y$-shape interaction has been suggested years ago~\cite{Artru:1974zn}, and often rediscovered either in models or in attempts to solve the QCD in the strong coupling limit~\cite{Dosch:1975gf,Gromes:1979xn,Hasenfratz:1980ka,FabreDeLaRipelle:1988zr,Carlson:1982xi,Bagan:1985zn}. This interaction is now confirmed by lattice QCD  \cite{Takahashi:2000te} (for further refs., see., e.g., \cite{Rothe:827579}).  It reads
\begin{equation}\label{eq:VY}
V_Y=\sigma\,\min_J\sum_{i=1}^3 r_{iJ}~,
\end{equation}
where each quark is linked to a junction $J$ whose location is optimized, as in the famous problem of Fermat and Torricelli. See Fig.~\ref{fig:Y}. Instead, the rule (\ref{eq:col-add}) with $v(r)=\sigma\,r$ would give a potential
\begin{equation}\label{eq:Vhalf}
V_3=\frac{\sigma}{2}\left( r_{12}+r_{23} +r_{31}\right)~,
\end{equation}
which is smaller but very close to $V_Y$, so that the phenomenology hardly distinguishes between the two potentials $V_Y$ and $V_3$~\cite{Richard:1983mu}.

In the case of tetraquarks (two quarks and two antiquarks), however, it was shown that the color-additive rule (\ref{eq:col-add}) and the generalization of the $Y$-shape potential lead to rather different spectra. In the former case, the stability of the tetraquarks $(qq\bar{Q}\bar{Q})$ requires a large mass ratio for the quarks and antiquarks. The latter potential, if alone and acting without any antisymmetrization constraint (e.g., with quarks and antiquarks of different flavors), gives stable tetraquarks~\cite{Vijande:2007ix}.  The tetraquark potential was taken as the minimum energy of two separate quark--antiquark flux tubes (the so-called ``flip-flop'' interaction) and a connected double-$Y$ Steiner tree linking the quarks to the antiquarks, a model of confinement which is supported by lattice QCD \cite{Okiharu:2004ve}.

Our aim is to extend the study of Vijande et al.\ to pentaquark states. For simplicity, we assume that the constituents have the same mass, but remain distinguishable  through their spin and flavor degrees of freedom, so that the orbital wave function can contain a symmetric component. We also consider the case where one of the constituents is infinitely massive. This paper is organized as follows. The model is described in Sec.~\ref{se:model}. The results are shown in Sec.~\ref{se:resu}, and some further investigations are suggested in Sec.~\ref{se:out}.
\section{The model}\label{se:model}
We focus on the role of confining forces, and hence disregard the Coulomb-like contributions and spin-dependent forces. For $(\bar{q}q)$ mesons, the Hamiltonian of the relative motion reads
\begin{equation}\label{eq:Hmes}
H_2=\frac{\vec{p}^2}{m}+\sigma\, r~,
\end{equation}
where  $\vec{p}$ is conjugate to the quark--antiquark separation $\vec{r}$, $r=|\vec{r}|$, and $m$ is  the constituent mass. For this system and the ones considered below, it is possible to set $m=\sigma=1$ without loss of generality, since departing from these values results in a simple scale factor  $m^{-1/3}\sigma^{2/3}$ of all the eigenvalues. The ground-state $(\bar{q}q)$ of (\ref{eq:Hmes}) can be expressed in terms of the Airy function, and its energy is  $E_2\simeq 2.33811$. For $(\overline{Q}q)$ or $(\bar{q}Q)$ with a static quark or antiquark, and a constituent of mass $m=1$, the reduced mass is twice larger, and by scaling the ground-state energy is $E_2'\simeq1.8558$

\begin{figure}
\begin{center}
\includegraphics[width=.50\textwidth]{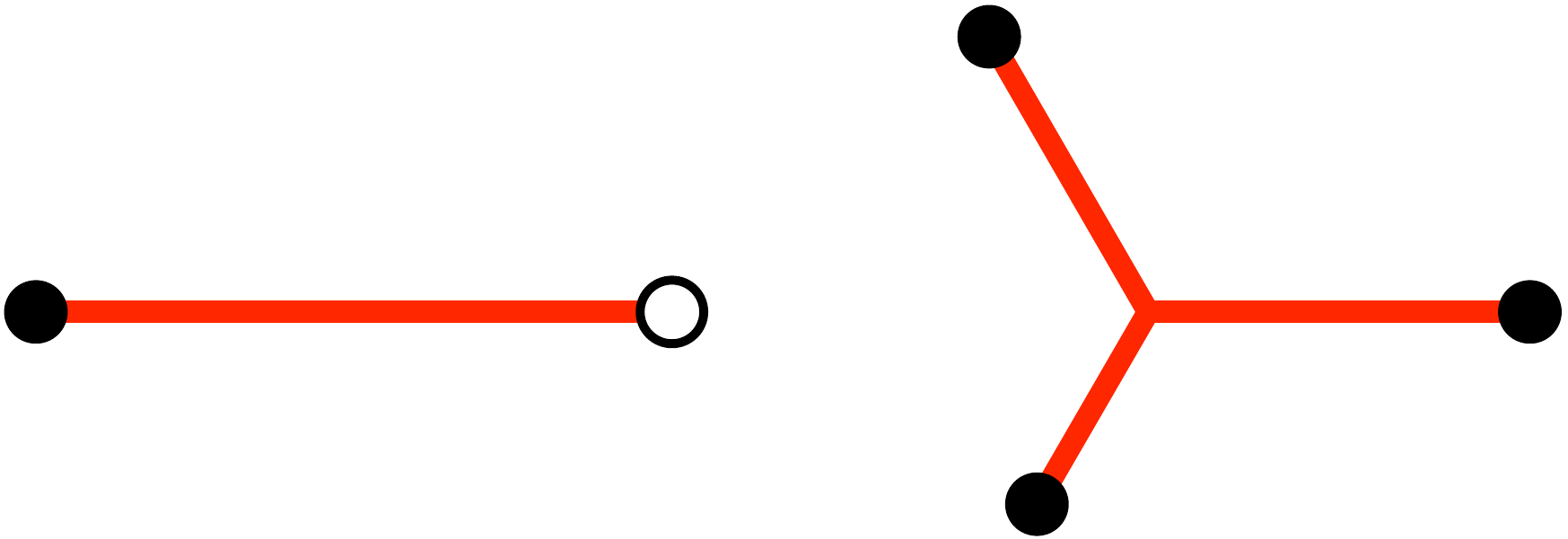}
\end{center}
\caption{\label{fig:Y} (Coloer on line) Confinement of mesons and baryons. The minimum over the quark permutations gives the flip--flop potential.}
\end{figure}

For $(qqq)$  baryons, we consider first the additive model (\ref{eq:Vhalf}), 
\begin{equation}\label{eq:H3}
H_3=\vec{p}_x^2+\vec{p}_y^2+\frac{1}{2}\left( r_{12}+r_{23} +r_{31}\right)~,
\end{equation}
where $\vec{x}=\vec{r}_2-\vec{r}_1$ and $\vec{y}=(2\vec{r}_3-\vec{r}_2-\vec{r}_1)/\sqrt{3}$ are Jacobi variables suited for equal masses, and $\vec{p}_x$ and $\vec{p}_y$ their conjugate momenta. We also study the more realistic $Y$-shape interaction (\ref{eq:VY}), schematically pictured in Fig.~\ref{fig:Y}, with the Hamiltonian
\begin{equation}\label{eq:HY}
H_Y=\vec{p}_x^2+\vec{p}_y^2+V_Y(\vec{r}_1,\vec{r}_2,\vec{r}_3)~.
\end{equation}
For $(Qqq)$, the kinetic-energy part is replaced by $(\vec{p}_1^2+\vec{p}_2^2)/2$.

For the five-body problem with equal masses, we start, as in~\cite{Paris:2005sv}, from a symmetrized model
\begin{equation}\label{eq:H5}
H_5=\sum_{i=1}^4 \vec{p}_i^2+\frac{1}{4} \,\sum_{i<j} r_{ij}~,
\end{equation}
where the cumulated strength encountered in a meson and in a baryon within the additive model is spread over the 10 interacting pairs.  Here, the relative motion is described by four Jacobi variables $\vec{x}_i$, to be specified shortly, and the $\vec{p}_i$ are their conjugate momenta.  Up to an irrelevant factor, the potential in (\ref{eq:H5}) is identical to the confining interaction adopted in Ref.~\cite{Genovese:1997tm}.

Note that, in the additive model,  the threshold can be understood as another five-body problem,
\begin{equation}\label{eq:Hth}
H_\text{th}=\sum_{i=1}^4 \vec{p}_i^2+ (1-\epsilon)r_{12}+\left[ \frac{1}{2}-\epsilon\right]\sum_{3\le i<j} r_{ij}+ \frac{2\,\epsilon}{3}\sum_{3\le i} (r_{1i}+r_{2i})~,
\end{equation}
in the limit where $\epsilon\to 0$. Then the variational principle applied to $H_\text{th}$ with the symmetric ground-state of $H_5$ as trial function immediately indicates that the lowest energy of $H_5$ is \emph{above} the threshold. In other words, in the color-additive model (\ref{eq:col-add}), the most asymmetric distribution of couplings is encountered in a threshold made of two separate color singlets, and this asymmetry benefits  the threshold and penalizes tentative multiquarks. This is confirmed by the results of Hiyama et al.~\cite{Hiyama:2005cf,PTPS.168.101}. To build stable multiquarks, one can either introduce a competing asymmetry by using different constituent masses, as done for the tetraquark $(QQ\bar{q}\bar{q})$,  or modify the color-additive model~\cite{Vijande:2009xx}.

\begin{figure}
\begin{center}
\includegraphics[width=.3\textwidth]{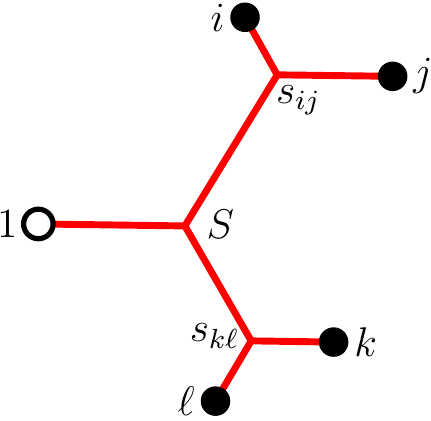}
\end{center}
\caption{\label{fig:Steiner5} (Color on line) Example of connected Steiner tree (planar for simplicity) linking the antiquark $1$ to the quarks $\{i,j,l,\ell\}$. The Steiner points are $S$, $s_{ij}$ and $s_{k\ell}$.}
\end{figure}

For the pentaquark, we consider the natural extension of the minimal-path model already used for the tetraquark~\cite{Carlson:1991zt,Vijande:2007ix,Ay:2009zp} and supported by lattice studies \cite{Okiharu:2004ve}.  It reads
\begin{equation}\label{eq:VP}
H_P=\sum_{i=1}^4 \vec{p}_i^2+V_P~,\qquad V_P=\min(V_\text{ff},V_\text{St})~,
\end{equation}
where
\begin{equation}\label{eq:Vff}
V_\text{ff}=\min_i\left[r_{1i}+ V_Y(\vec{r}_j,\vec{r}_k,\vec{r}_\ell)\right]~,
\end{equation}
with $\{i,j,k,\ell \}$ being any permutation of $\{2,3,4,5\}$, is the so-called flip--flop potential. It corresponds to the most economical configuration in Fig.~\ref{fig:Y}.  The second term, $V_\text{St}$, corresponds to a connected Steiner tree, as pictured in Fig.~\ref{fig:Steiner5}.  
This Steiner tree generalizes the Fermat--Torricelli problem for more than three points. The quarks $(i,j)$ are linked to the central Steiner point $S$ as three quarks in an ordinary baryon, through an intermediate Steiner point $s_{ij}$. Similarly, $(S,k,\ell)$ form a baryon-like structure with Steiner point $s_{kl}$. Then, the antiquark and the two intermediate Steiner points $s_{ij}$ and $s_{kl}$ form an antibaryon-like configuration. Of course, the potential $V_\text{St}$ is optimized by varying the permutation $\{i,j,k,\ell \}$ of the quarks. See, e.g., \cite{Bicudo:2008yr,Ay:2009zp} for references on the Steiner problem and its application to the multiquark potential. 

Note that the long-range part of the pentaquark potential, as estimated in lattice QCD, has been found fully compatible with this multi-$Y$ term \cite{Okiharu:2004wy}.

An obvious consequence of the minimization of (\ref{eq:VP}) is that 
\begin{equation}\label{eq:ineg}
V_P\le r_{12}+ V_Y(\vec{r}_3,\vec{r}_4,\vec{r}_5)~,
\end{equation}
indicating that the pentaquark potential is smaller than the cumulated confinement energy of the threshold, i.e.,  that the effective potential between the meson $(1,2)$ and the baryon $(3,4,5)$ is attractive. However, in  three space dimensions, an attractive potential does not automatically lead to binding, if this potential is short-ranged. Hence one should solve the five-body problem to determine whether this model supports or not a stable pentaquark.

In the case of the tetraquark, it was noticed~\cite{Vijande:2007ix} that the connected Steiner tree configuration, though the most interesting, plays a minor role, and that the binding is obtained from the flip--flop term alone. Similarly, a survey with randomly generated coordinates for the antiquark and the four quarks shows that, in  Eq.~(\ref{eq:VP}), the Steiner tree gives the minimum in less than $3\%$ of the cases. Hence, we shall neglect this term and compute an upper bound with the flip-flop interaction alone. This is opposite to the choice made in \cite{Paris:2005sv}, where only the connected-Steiner-tree term is adopted and the flip--flop one omitted.
\section{Results}\label{se:resu}
The ground-state meson can be described by a simple expansion
\begin{equation}\label{eq:gauss2}
\Psi_2=\sum_i\gamma_i\,\exp(-\alpha_i \,r^2/2)~,
\end{equation}
or, in short, $|\Psi_2\rangle=\sum_i\gamma_i\,|\alpha_i\rangle$,  and  the energy $E_2\simeq 2.33811$ can be reproduced with just a few terms. The relevant matrix elements $\langle\alpha' |\alpha\rangle$, $\langle\alpha' |\vec{p}^2|\alpha\rangle$ and $\langle\alpha' |r|\alpha\rangle$ are known analytically, and are basic ingredients for more complicated systems.

For the ground-state of $H_3$, a generalization reads
\begin{equation}\label{eq:gauss3}
\Psi_3=\sum_i\gamma_i\,\exp[-(a_i \,\vec{x}^2+b_i\,\vec{y}^2+ 2 c_i\,\vec{x}.\vec{y})/2]~,
\end{equation}
and, again, the matrix elements are known analytically. For a given choice of range parameters, the weights $\gamma_i$  are given by a generalized eigenvalue equation. To avoid ambiguities and simplify the minimization, one can restrict the Gaussians to scalar ($a=b$, $c=0$) or diagonal ($a\neq b$, $c=0$) matrices, and those given by permutation of the quarks; furthermore the parameters $a_i$ and $b_i$ can be taken from a single set $\{\alpha, \alpha+\delta, \alpha+2\delta, \ldots\}$, with minimization only over the two extreme values~\cite{2003PrPNP..51..223H}. One  reaches $E_3\simeq 3.863$. If only scalar matrices are allowed, the expansion (\ref{eq:gauss3}) converges towards the best function  of the hyperradius given by $\rho^2=\vec{x}^2+\vec{y}^2$. In this approximation, 
$u(\rho)=\rho^{5/2}\,\Psi$  is given by the radial equation
\begin{equation}\label{eq:hyper3}
-u''(\rho)+\frac{15}{4\rho^2}u(\rho)+V_{00}\,\rho\, u(\rho)=E_3^{(0)}\,u(\rho)~,
\end{equation}
with suitable boundary conditions, which leads to the upper bound  $E_3^{(0)}\simeq 3.865 $. The hyperscalar projection, including the strength factor $1/2$ and the number of pairs, is 
\begin{equation}\label{eq:hypercoff3}
V_{00}=\frac{16}{5\pi}\simeq1.019~.
\end{equation}

A similar strategy can be used for the $Y$-shape potential, except that the matrix elements have to be calculated numerically. The hyperscalar coefficient becomes $V_{00}= 1.115$. By scaling from (\ref{eq:hypercoff3}), this corresponds to an energy $E_Y^{(0)}\simeq 4.105$.  The Gaussian expansion, if not restricted to scalar matrices, gives a better energy $E_Y\simeq 4.095$. This means that, in our simple string model, the threshold for the stability of light pentaquark states is
\begin{equation}\label{eq:threshold}
E_\text{th}(\bar{q}qqqq)\simeq 6.433~.
\end{equation}
If a static quark $Q$ or antiquark $\overline{Q}$ is introduced, 
for  $(\overline{Q}qqqq)$, the threshold consists of $(\overline{Q}q)+ (qqq)$, and for $(\bar{q}qqqQ)$, it is the lowest of $(\bar{q}Q)+ (qqq)$ and $(\bar{q}q)+ (Qqq)$, which turns out to be the latter by a small margin, and the thresholds are
\begin{equation}\label{eq:th-Qqqqq}
E_\text{th}(\overline{Q}qqqq)\simeq 5.950~,
\qquad
E_\text{th}(\bar{q}qqqQ)\simeq 5.944~.
\end{equation}
We now turn to the five-body problem, first for $(\bar{q}qqqq)$. A possible choice of Jacobi variables, besides the center-of-mass, is  
\begin{equation}\label{eq:Jacobi5}
\begin{aligned}
\vec{x}_1&=\frac{4\vec{r}_1-(\vec{r}_2+\vec{r}_3+\vec{r}_4+\vec{r}_5)}{\sqrt{10}}~,
\quad
&\vec{x}_2&=\frac{\vec{r}_2-\vec{r}_3+\vec{r}_4+\vec{r}_5}{2}~,\\
\vec{x}_3&=\frac{\vec{r}_2+\vec{r}_3-\vec{r}_4+\vec{r}_5}{2}~,
&\vec{x}_4&=\frac{\vec{r}_2+\vec{r}_3+\vec{r}_4-\vec{r}_5}{2}~,
\end{aligned}
\end{equation}
which is convenient to express the permutations of the quarks. With a Gaussian expansion, one obtains for the symmetric toy model $H_5$ an energy $E_5\simeq 6.850$. In the hyperscalar approximation, the radial equation is similar to (\ref{eq:hyper3}) with now a centrifugal coefficient
$99/4$ instead of 15/4, and a strength $V_{00}=2560/(693\pi)\simeq1.176$, and a very similar energy is found. This model $H_5$ has thus energies well above the threshold $E_2+E_3$ that includes pairwise forces for the baryon, and even above the threshold $E_\text{th}=E_2+E_Y$ when the baryon is bound by the $Y$-shape interaction.

If now one switches to the flip--flop interaction, the coefficient  of the  hyperscalar potential 
$V_{00}\,\rho$ is found to be $V_{00}=1. 031$, by numerical integration of $V_\text{ff}$ over the hyperscalar variables, for given $\rho$. By scaling, this gives an energy 
\begin{equation}\label{eq:hyper5ff}
E_\text{ff}^{(0)}(\bar{q}qqqq)\simeq 6.276~,
\end{equation}
in the hyperscalar approximation, which is clearly \emph{below} the dissociation threshold (\ref{eq:threshold}). This is confirmed by the Gaussian expansion. As these two methods are variational, the flip-flop model gives a bound state, and this is \textsl{a fortiori} the case if the connected Steiner trees are also included in evaluating the potential (\ref{eq:VP}).

If the calculation is repeated in the case of four unit masses and an infinitely massive antiquark or quark,  the kinetic energy in (\ref{eq:VP}) is replaced by $\sum \vec{p}_i^2/2$, where $\vec{p}_i$ is conjugate to the position $\vec{r}_i$ of a finite-mass constituent. 
The pairwise model (\ref{eq:H5}) gives an energy of  about 6.305 for both $(\overline{Q}qqqq)$ and $(\bar{q}qqqQ)$ in the hyperscalar approximation, i.e., well above the threshold, and also with the Gaussian expansion, no state is found below the threshold. 
On the other hand, the flip--flop potential, treated in the hyperscalar approximation, gives
\begin{equation}\label{eq:ff-Qqqqq}
E_\text{ff}^{(0)}(\overline{Q}qqqq)\simeq 5.836~,
\quad
E_\text{ff}^{(0)}(\bar{q}qqqQ)\simeq 5.667~,
\end{equation}
which is sufficient to demonstrate the of  stability of $(\overline{Q}qqqq)$ and $(\bar{q}qqqQ)$ with respect to their respective thresholds~(\ref{eq:th-Qqqqq}).
\section{Conclusions and outlook}\label{se:out}
A simple string model  of linear confinement gives a pentaquark which is stable against spontaneous dissociation into a meson and an isolated baryon. This is at variance with most  of the earlier constituent-model calculations.  There are, however, severe limitations in our approach:
\begin{itemize}
\item
The non-relativistic kinematics for the quarks  and the Born--Oppenheimer treatment of the gluon field, which supposedly readjusts itself immediately when the constituent move, call for an application to heavy quarks. However, the short-range central corrections should be incorporated for heavy quarks. For the Coulomb-like interaction, the color-additive rule (\ref{eq:col-add}) probably holds. If alone, a Coulomb interaction with color factors will not bind. Our pure confining model binds. What occurs for a superposition would be a matter of  detailed phenomenology beyond the scope of the present note. 
\item
The quark wave function is assumed to be compatible with an overall s-wave, and is symmetric under permutations. Thus, it should be associated with enough spin and flavor degrees of freedom in the quark sector. A proper treatment of the Fermi statistics of quarks is rather delicate in the flip--flop or Steiner-tree model, as different flux tube topologies correspond to different color couplings, and thus different constraints for the spin, flavor and space parts of the wave-function.
\item
The calculation has been restricted to five equal and finite masses, or one infinitely massive constituent surrounded by four equal masses. The property of stability was found to survive for both $(\overline{Q}qqqq)$ and $(\bar{q}qqqQ)$. The investigation should be extended with better variational wave functions, and a larger variety of mass distributions for the constituents. Clearly the configuration $(\bar{c} uuds)$ and its analogs by permuting the light quarks or by replacing $c$ by $b$ would deserve a refined treatment  including the spin--spin forces, which are favorable \cite{Lipkin:1987sk,Gignoux:1987cn}.
\item
The pentaquark states with two or more heavy constituents would deserve a specific studies. For $(QQq)$ baryons, the dynamics can be studied in the Born--Oppenheimer approximation, where the two heavy quarks experience an effective interaction resulting from their direct interaction modified by the light quark \cite{Fleck:1989mb,Yamamoto:2007nn,Najjar:2009da}. Similarly, the $(QQ\bar{q}\bar{q})$ tetraquarks and the $(QQqq\bar{q})$ or $(qqqQ\bar{Q})$ pentaquarks could be studied through in the adiabatic limit.
\end{itemize}

It remains that our conclusion, based on the flip--flop dynamics, is drastically different from the one obtained from the connected Steiner tree alone \cite{Paris:2005sv}. Studies within the lattice QCD \cite{Okiharu:2004ve} or, more recently, the AdS/QCD~\cite{Andreev:2008tv}, have analyzed the interplay between flip--flop and connected multi-$Y$ configurations for tetraquarks. It would be desirable to have more information about the analogue for pentaquarks.

\acknowledgments{It is a pleasure to thank M.~Asghar and S.~Fleck for fruitful discussions.}

%
%

\end{document}